\documentclass[12]{article}
\usepackage{mathtext}
\newcommand{\nn}{\medskip} 
 
\renewcommand{\Im}{{\rm Im}}
\renewcommand{\d}{\delta}
\renewcommand{\b}{\beta}
\newcommand{\g}{\gamma}

\newcommand{\e}{\epsilon}

\newcommand{\ar}{\longrightarrow}
\newcommand{\w}{\omega}
\newcommand{\s}{\sigma}
\newcommand{\la}{\lambda}
\renewcommand{\a}{\alpha}
\begin{document}
\title{Simulation of quantum interference by reactions of chemical type\thanks{Russian version of this paper is available at http://qi.cs.msu.su}}
\author{Yuri I. Ozhigov\thanks{e-mail: 
ozhigov@ftian.oivta.ru.} \\[7mm]
Moscow State University,\\
Institute of Physics and Technology,\\
Russian Academy of Sciences,\\ 
} 
\maketitle
\begin{abstract}A quantum unitary evolution alternated with measurements is simulated by a bubble filled with fictitious particles called amplitude quanta that move chaotically and can be transformed by the simple rules that look like chemical reactions. A basic state of simulated system is treated as a collision of the two corresponding amplitude quanta, that gives the quantum statistics of measurements. The movement of the external membrane of the bubble corresponds to the classical dynamics of the simulated system. Measurements are treated as the membrane perforations and they are completely determined by initial conditions. An identity of particles and an entanglement is simulated by the membranes touching. 
The required memory grows linearly where the number of particles increases, but entangled states of the big number of particles can be simulated. The method can be used for a visualization of quantum dynamics. 

\end{abstract}

\section{Outlook of the method}

The problem of reorganization of quantum mechanics in order to obtain the theory more effective in the sense of predictive power or computational complexity, is one of the most intriguing scientific problems (look at \cite{Pe, Be, Kh}). Nevertheless the single conventional mathematical formalism of Hilbert spaces is so general and completed that attempts to gain such a reorganization in the framework of this formalism seems to be not very promising. 

In this paper we touch the more practical aspect of this general problem: the possibilities of simulation and visualization of quantum evolutions by classical computers. 
We do not try to give an effective description of quantum mechanics in terms of classical algorithms - it would be wittingly unrealizable task in view of peculiar quantum phenomena like quantum speedup of algorithms, based on the huge dimensionality of Hilbert spaces (see \cite{Sh, Gr}), quantum communications (see \cite{Ho, Ben}), and the relativistic features of space-time which require the consideration of unlimited intervals in the space and time (see \cite{Mo}).  

Instead of it we take up the creation of physically right visual "picture" of events happening on the quantum level. Such animated "picture" must show us what we could in principle see if we overcome the technological difficulties\footnote{Lev Landau said that in quantum physics we can understand some things which we cannot imagine. Our imagination is brought up mainly in the classical spirit and it lags behind the possibilities of the mathematical formalism of quantum physics. Nowadays this weakness of imagination becomes the serious obstacle on the way of effective simulation of atomic and molecular processes.}. Beyond the "picture" visible to a user there must be an invisible part in the model which has no physical sense but ensures the right functioning of the "picture", that includes an agreement with quantum laws and an easy scalability, e.g., a possibility to include new particles and interactions - in the framework of the possibilities of the classical memory.  

Such a visualization of events going in the spatial areas of a few angstroms size is very important for the applications of quantum physics. For example, the right understanding of the main biochemical processes is impossible without taking into account the quantum effects:  rebuilding of electron states in chemical reactions, proton tunneling, interaction with electromagnetic fields. But the most important type of the quantum behavior that cannot be easily considered in the classical or quasiclassical simulation of molecules and elementary particles is their entaglement. Just entangled states represent the main problem in the classical simulation because the right consideration of such states requires the Hilbert space of exponential dimensionality (from the number of particles). We have every reason to beleive that such states play the significant role in functioning of nanostructures, so that the right picture cannot be created without a visual representation of entanglement. 

The simplest way for the visualization of one particle quantum evolution is as follows. We separate the minimal spatial area where this particle is localized with the probability close to one, and call this area the bubble corresponding to this particle. Its covering is called a membrane. If we look at this bubble from outside then the movement of its membrane must correspond to that we could in principle see if we use as strong magnification as possible. If we neglect the size of the bubble and take into account its "average" position in space, then this averaging of its movement must correspond to the classical dynamics accordingly to Erenfest theorem. The internal area of the bubble must be hidden from observations - just there the quantum evolution goes. It is then reasonable to treat a measurement as a perforation of the membrane that results in its rebuilding. A perforation must happen, for example, when the continuation of its movement immediately leads to its decomposition into two or more disconnected parts. Entangled states must arise when two membranes of the different particles touch each other. An expansion of such a state on some orthonormal basis  $|e_0,\rangle,|e_1\rangle,\ldots,|e_{N-1}\rangle$ has the form $\sum\limits_j\la_j|e_j\rangle$. If these states presume the spatial localization of all particles, then this superposition can be visualized if we represent in turn all the states $|e_j\rangle$, each in the short time $\d t|\la_j|^2$ proportional to the corresponding probability.

How can we represent what happens inside a bubble ?
We can simply solve the corresponding Shroedinger equation by any of the known numerical methods. The disadvantage of this direct way is that it requires enormous computational resources: the dimensionality of the Hilbert space of states grows exponentially when the number of particles grows. 
Here it often happens that this generality is redundant and a consideration in a much lesser space gives a good agreement with experiments\footnote{Of course, we should at once refuse from the attempts to simulate such hypotetic devices as quantum computers. In such devices just the huge dimensionality of the space of states plays the key role. In other words quantum computers presume the possibility for amplitudes to take infinitesimal values, whereas in the simulation we have to exclude this possibility, introducing the grains (the minimal positive values) for all magnitudes including amplitudes.}.
The choice of approximation then turns out to depend on a physicist intuition, that decreases the effectiveness of an algorithmic simulation. This is why we try to reduce a quantum interference to some classical procedure which is more appropriate for a numerical simulation. With such a procedure we hope to separate the small part of the huge Hilbert space of states - the part which is sufficient for the right simulation and visualization of quantum processes (look at simulation of entanglement).

The key difficulty here lies in the nature of quantum interference: it does not allow the natural dynamical description \footnote{The second difficulty is how to simulate a wave function collapse. A partial answer - when this collapse happens, was given in the paper \cite{Oz}. It happens when the current quantum state description requires more memory than we can reserve for it.}. This difficulty can be got round if we replace the probabilistic interpretation of amplitudes by an explicit urn scheme. We obtain finally a dynamical description of an interference, where the probability interpretation of amplitudes arises directly from the laws of probability applied to a classical chaotic behavior of particles in a limited reservoir. Here the measurement statistics is stable relative to the small changes of  initial conditions in our model, as well as for initial conditions of Shroedinger equation. The difference is that in our model an initial condition completely determines the results of all measurements and we can manage without an artificial generation of pseudo-random numbers during the simulation, all the randomness is concluded in an initial state of the model. This description of quantum interference can be thus called dynamical but in the sense of the dynamics of fictitious particles which are introduced in order to obtain such a description\footnote{It means practically, for example, that our approach allows a simple parallelization resistant to local hangs-up. If we directly parallelize the numerical solution of Shroedinger equation which Hamiltonian depends on the time $t$ then no local control can remove its vulnerability to local hangs-up. In our approach, on the contrary, an influence of any hang-up is limited by the fraction of hung processors.}. 

Our main aim is to build a classical model embracing a state vector $|\Psi\rangle$ with its probability interpretation, its unitary evolution corresponding to some Hamiltonian $H$, of the form 
$$
|\Psi(t)\rangle=T\{\exp [-\frac{i}{h}\int\limits_{t_0}^tH(t)dt]\}\ |\Psi(t_0)\rangle,
\footnote{$T$ is the chronological exponential operator.}$$
 and all naturally arising sequential measurements of this state. The redistribution of the scarce classical memory must be organized more rationally than in the conventional model, so that we could trace the sequentional transformations of entangled states of the form (\ref{entangled}) where the sum is expanded to the less than exponential number of states\footnote{Just such states are observed in experiments on interference of big molecules. Without the last limitation no visualization of quantum evolution is possible because we fall within the area of quantum computations.}. 

For this aim we introduce fictitious particles called amplitude quanta (a.q.) and its reactions similar to chemical reactions. 
We use the easy heuristic. Any amplitude can be approximately represented as a sum of numerous small numbers of the forms $g,-g,ig,-ig$, where $i=\sqrt{-1}$. These numbers are thought of as a contents of ficticious particles - a.q., that moves chaotically inside of some bubble ${\cal B}$ which border $\partial{\cal B}$ is thought of as the two dimensional membrane consisting of pointwise cells and it reflects all flying quanta. Neither a.q. no its dynamics has a physical sense, it serve as an alternative of the conventional approach (solution of Shroedinger equation and the following generation of random numbers with the obtained distribution) only. We can thus choose any form of a.q. movement and their interaction with the membrane proceeding from the conveniency of computations. We suppose that any amplitude quantum can react with the other closest quantum or with the cell of the membrane which this quanta collides with, where in the first case these two quanta do not influence to the movement of each other. We thus assume that the collision of quanta changes their types only but not the trajectories\footnote{In absentia of elastic collisions the trajectories become not so unstable as in the billiard model with elastic collisions between the particles: the difference between trajectories with the close initial conditions grows linearly with the time in opposite to the billiard case when it grows exponentially. The single source of divergence between trajectories is the collisions with the membrane that gives a weak instability only. This makes possible to introduce the options of simulation making this model deterministic and thus to manage without the generation of pseudo-random numbers: the complete information about the behavior of the model is contained in the initial conditions. It means that we can thus hope to get rid of the possible dependence on artifacts contained in pseudo-random generators.}. Our approach has one more adventage that is a simple representation of a decoherence (see simulation of classical dynamics by membrane movements). 

 A decoherence is usually treated as a result of vacuum fluctuations or the other chaotical influence of environment, e.g. it requires the introduction of a new random variable in addition to the main random variable which determines the results of measurements. In our approach a decoherence arises as a simple consequence of the bubble dynamics when the membrane is on the eve of the destruction. The only random variable in our model is the initial state of all a.q. and the membrane position. The "bubble" model even without a.q. requires less independent random variables than the conventional model. In addition, a.q. make possible to use the same program primitives for the simulation of the quantum and classical chaotic behavior.  

Given a basic state of the quantum ensemble the corresponding amplitude can be obtained if we sum as numbers all a.q. in the bubble corresponding to this state. Nevertheless, for the simulation of interference it is convenient sometimes to keep in the bubble quanta with the opposite signs, like $g$ and $-g$, $ig$, $-ig$; moreover, in some cases the difference between the quantities of quanta of the same type with the opposite signs must be much less than their total number. The physical space here is treated as filled by bubbles which can move, change the shape, merge and split up with the membrane rebuilding (in the measurement), and also they can touch by the membranes and exchange their a.q. (see the simulation of identical particles). 

\section{Bubble with amplitude quanta}

We describe a quantum one particle evolution by means of a bubble filled with chaotically moving fictitious free particles called amplitude quanta (a.q.) which interact mechanically with the membrane only and interact with each other and with the membrane through the reactions of chemical type. Consider a three dimensional bubble ${\cal B}$ with the two dimensional membrane $\partial{\cal B}$ which is divided into the non-overlapping areas 

\begin{equation}
\partial{\cal B}=\bigcup\limits_{j=1}^l\partial{\cal B}_j.
\label{border}
\end{equation}
 This bubble is filled by $m$ a.q. The different a.q. differ
by their spatial positions, speeds and by their types. A type of a.q. is a subject for change in its collisions with the others a.q. and with the membrane. A.q. are thought of as pointwise; it is convenient to assume that the different a.q. do not change their trajectories and speeds when they collide with each other, and the membrane reflects any a.q. elastically with the conservation of the angle, if the opposite is not stated explicitly. Given a precision of spatial location $r_0$ we assume that the collision happens if two a.q. or one a.q. and the membrane become closer than the minimal distance $r_0$. Laws for the change of types after the collision are called reactions. The shape of bubble is determined by its a.q., and the division (\ref{border}) is the subject for external control during the simulation\footnote{The more radical viewpoint is that all divisions of the membrane are induced by its contact with the membranes of the others bubbles, for example in the case of interactions between the considered particle and the electromagnetic field the divisions are induced by the membranes of photons.}. 

Given an amplitude quantum $q$ we denote its type by $\tau (q)$. Every type has the form 
\begin{equation}
x^s_r,
\label{quantum}
\end{equation}
 where $x\in \{\a,\b\}$ determines which part of amplitude is represented by this quantum: real ($\a$) or imaginary ($\b$), $s\in\{ +,-\}$ determines the sign of this quantum, and $r$ is the list of the form $r=j\ r_1\ \ldots \ r_k$. Here the first element $j=0,1,\ldots,N-1$ determines the basic state $|\Psi_j\rangle$ which the amplitude corresponds to (see below) and the rest elements contain the auxiliary options of the quantum. We assume the conventional rules for operations with signs. We denote by $[ x^s_j]_{\cal B}$ the total number of quanta  of the form (\ref{quantum}) in the bubble ${\cal B}$, where in the lower index the auxiliary options as well as ${\cal B}$ will be often omitted. We put $ [x_j] = [x^+_j]- [x^-_j]$.
The reactions between a.q. can have one of the following types. 
\begin{equation}
\begin{array}{llll}
&1). &\tau,\ M &\ar \tau',\ M' ,\\
&2). &x_r^s,\ x_r^{-s}&\ar \emptyset.\\
&3). &\tau_1,\ \tau &\ar \tau'_1,\ \tau .\\
&4). &\tau,\ M &\ar \tau,\ \tau',\ M.
\end{array}
\label{general_reactions}
\end{equation}
The type 1) represents a simple transformation of a.q. of the type $\tau$ when it collides with a membrane cell of the type $M$. The types of a.q. and a cell here change.  
The type 2) represents an annihilation of two corresponding a.q. with the opposite signs. 
The type 3) represents a "catalysis" with a reagent of the type $\tau_1$ and a "catalyst" of the type $\tau$. 
Here two a.q. of the types $\tau_1$ and $\tau$ collide and turn into the a.q. of the types $\tau'_1$ and $\tau$ correspondingly without changing their speeds. The "catalysis" thus influences on the type of reagent only but not on its dynamics. 
The type 4) represents a nonequilibrium reaction where the total number of a.q. is not conserved. The type of membrane cell here is conserved and we shall omit it.

To simulate a unitary evolution it is sufficient to use reactions of 
the form 4), or alternatively, reactions of the type 3). To simulate measurements we need reactions of the form 2) and 
reactions with a virtual state (see below). 

The membrane is the two dimensional border isolating the internal area of the bubble from the outside. It consists of pointwise cells of the different types. There is no more than one special cell in the membrane called a virtual state. There are three types of a virtual state:
$$
\begin{array}{lll}
&a)\ \ \ &(\ ,\ ),\\
&b)\ \ \ &(\tau(q),\ ),\\
&c)\ \ \ &(\tau(q),\tau(q)),
\end{array}
$$
that is it consists of two places for a.q., and these places can be occupied or free; if there is only one a.q. in the virtual state it occupies the first place. A virtual state of the type $c)$ is called a real state.
The reactions between a virtual state and a.q. can have the following forms:
\begin{equation}
\begin{array}{lll}
&(\ ,\ );\ \tau(q)&\ar (\tau(q),\ ),\\
&(\tau(q),\ ); \ \tau(q)&\ar (\tau(q),\tau(q)),\\
&(\tau(q),\ ); \ \tau(q')&\ar (\tau(q'),\ ),\tau(q)),\ \ (\tau(q)\neq\tau(q').\\
\end{array}
\label{virt}
\end{equation}
The first reaction means that every flying a.q. occupies the first place of the virtual state provided the both places are free. The second means that if the first place of the virtual state is occupied by some a.q., and the second is free, then the flying a.q. occupies the second place if and only if it has the same type. The third reaction means that if the flying a.q. has the different type, then it replaces the a.q. which was in the virtual state; that a.q. becomes free and follows the way of the flying a.q. as if the virtual state is the ordinary cell. How does the virtual state work? It "waits" for the situation when two a.q. of the same type collide to it sequentially, as it happens the virtual state becomes a real state.

In addition we can introduce 
the reactions between membrane cells of the form of cellular automaton. It may be used to simulate a quantum evolution with a Hamiltonian depending on the time $t$ or to initiate a measurement.

The result of all possible annihilations of a.q. of the type $x_r,\ x_r^{-1}$ is called $r$- reduction. 
Given a bubble with a.q. we define for it the real nonnegative numbers
\begin{equation}
p_j=\frac{ [\a_j]^2+[\b_j]^2}{ \sum\limits_{x\in A,\\ 0\leq k\leq N-1}[x_k]^2}.
\end{equation}
Such a number $p_j$ can be considered as a probability to obtain a real state of the type $(x^s_j,x^s_j)$ for any $x\in A,\ s\in\{ +,-\}$ as a result of all sequential $j$- reductions in the bubble ($j=0,1,\ldots ,N-1$) and the following reactions of the form (\ref{virt}), provided we start from the virtual state of the type $(\ ,\ )$ and perform no reactions between a.q. It makes possible to simulate a measurement procedure by the reactions between a virtual state and a.q.

A membrane plays the central role in our approach. It isolates the internal space of the bubble from the outside and thus forms the appropriate media for the chaotic movement of a.q. inside the bubble. It controls over the model of unitary evolution through the inheritance of the membrane cells types by the flying a.q. It controls over the model of measurement. 
It could be also used for the organization of reactions between a.q. if its cells are built as a virtual state.
It serves for placing of a key object of the simulation - a virtual state. But the main function of a membane is to isolate the space open for physical observation - the external space from the internal area of a bubble which is inaccessible for such observations in principle. The processes, going inside the bubble, have no physical sense; they are needed exclusively to ensure the right functioning of the model, where a user can see the external part only. We thus must not look after the observance of the physical laws for a.q., the conventional assumptions like "free will" are invalid for them - it can exist outside bubbles only. A.q. must not obey the fundamental limitations following from this assumption, for example the relativistic limitation on the speed of information transfer - there is no information without the "free will". The transmutations of a.q. must ensure the observance of the physical laws outside the bubbles, where the experiments are possible. In this paper we show how to ensure the observance of the quantum law of amplitudes interference in the framework of this approach. \footnote{It is very likely that the wide class of phenomena can be simulated in the framework of the "bubble approach": photoeffect, simple chemical ractions, EPR - pairs of photons, etc.}

\section{Representation of one particle states by amplitude quanta}

In this section we introduce the notions that form the "chemical" model of a quantum evolution. At first we describe these notions for a system with the constant Hamiltonian $H$ and basic states $\Psi_0,\Psi_1 ,\ldots,\Psi_{N-1}$ which must not be eigenvectors of $H$. We assume that all magnitudes represented in the model are grained.\footnote{In case of classical colliding dynamics this assumption would immediately leads to the strong form of irreversibility, but in our case we exclude elastic collisions and the assumption about the grained magnitudes is consistent with the reversibility in the short time frame.} The important consequence from this assumption is that amplitudes of all quantum states must be grained, e.g. must have the form $\a +i\b$, where $\a ,\b\in\{ 0,g,-g,2g,-2g,\ldots\}$ for some fixed grain $g>0$. It makes possible in the framework of our model to replace the probability interpretation of an amplitude squared module by the explicitly defined urn scheme. That is the probabilities of some favorable outcome will be simply its frequency in numerous iterations of the experiment given by the formula $p=N_{fav.}/N_{total}$. 
Given a complex number $a+ib$ we denote by $\a,\b$ its real and imaginary parts correspondingly, and put $A=\{ \a,\b\}$. 

Given a state 
\begin{equation}
|\Psi\rangle=\sum\limits_{j=0}^{N-1}\la_j|\Psi_j\rangle
\label{Psi}
\end{equation}
 of the particle we denote ${\rm Re}\ \la_j,\ \Im\ \la_j$ by $\a_{j,\Psi},\ \b_{j,\Psi}$ where the lower index $\Psi$ will be often omitted. We represent a normalized state (\ref{Psi}) by some bubble ${\cal B}$.

The state (\ref{Psi}) is called corresponding to the bubble ${\cal B}$ 
if and only if for all $j,k=0,1,\ldots,N-1$ and $x,y\in A$ the following equations are true:
$$
\frac{[x_j]_{\cal B}}{[y_k]_{\cal B}}=\frac{x_{j,\Psi}}{y_{k,\Psi}}.
$$
In this case we write $|\Psi\rangle=|\Psi\rangle_{\cal B}$. Applying the conventional rules for the calculation of the probabilities we conclude that if $\Psi=\Psi_{\cal B}$ then for all $j=0,1,\ldots,N-1$ $\ p_j=|\la_j|^2$ that substantiates the probability interpretation of the squared module of amplitude. We thus reduce this interpretation to the classical urn scheme by means of the reactions of two types: reductions and (\ref{virt}).   
The state of the bubble ${\cal B}$ in time instant $t$ is denoted by ${\cal B}(t)$. 

We now consider the time dependence of the state $|\Psi\rangle_{\cal B}$. We put ${\cal B}={\cal B}(0)$ and let $L$ be a list of reactions of the form (\ref{general_reactions}), $t>0$ be time segment. We denote by $L_{\cal B}(t)|\Psi\rangle$ the state $|\Psi\rangle_{{\cal B}(t)}$, where $|\Psi\rangle=|\Psi\rangle_{\cal B}$. For a random choice of initial conditions of amplitude quanta in the bubble and the big total number of them this state can be considered as independent from ${\cal B}$. In what follows we neglect the small influence of the bubble shape and its a.q. initial conditions to all our considerations. The correspondence between a state vector unitary evolution $U(t)$ and a list $L$ of reactions can thus be expressed by the following equation
\begin{equation}
U(t)=L_{\cal B}(t)
\label{bubble}
\end{equation}
which accuracy determines the fidelity of the bubble model. A list $L$ is called valid if the transformation given by (\ref{bubble}) is unitary. Two valid lists of reactions $L$ and $L'$ are called equivalent relative to a state $|\Psi\rangle$ if for any bubble ${\cal B}$ and time instant $t$ the following equation is satisfied $L_{\cal B}(t)|\Psi\rangle= L'_{\cal B}(t)|\Psi\rangle$. 

The only thing we should define to determine the model of one particle unitary evolution is the list of reactions between a.q. In this paragraph we assume that the membrane does not contain a virtual state and there are no $j$- reductions for a.q. inside the bubble. We show how to build this list for the basic unitary one qubit transformations induced by the action in time frame $\delta t$ of the Hamiltonians $H$, represented by the Pauli matrices taken with the opposite sign $-\sigma_x,\ -\sigma_y,\ -\sigma_z$. These transformations always have the form $\exp(-iH\delta t)$. At first we shall consider the action of Hamiltonians in the short time frame $\delta t$, but then it will be established that the obtained solution remains true for the times $O(1)$. For the short time frame the corresponding unitary transformation can be represented as $U(\delta t)=1+iH\delta t$ within $O((\delta t)^2)$ and we make all calculations with this accuracy. 

\subsection{Nonequilibrium reactions simulating Hamiltonians in the form of secondary quantization} 

We can build the list of nonequilibrium reactions easily for Hamiltonians written in the form of secondary quantization: through the operators of creation $a_j^+$  and annihilation $a_j$ of a particle in a state $j$. At first we consider the case $H=-\s_x=-\left(\begin{array}{lll}
&0&1\\
&1&0
\end{array}
\right)$.  
\nn

This Hamiltonian can be represented as follows:
$$
H=-a_2^+a_1-a_1^+a_2.
$$
Shroedinger equation $\frac{\partial}{\partial t}|\Psi\rangle=-iH|\Psi\rangle$ gives the following equations for the real and imaginary amplitude components:
$$
\begin{array}{ccc}
&\frac{d}{dt}\a_1&=-\b_2,\\
&\frac{d}{dt}\b_2&=\a_1,\\
&\frac{d}{dt}\a_2&=-\b_1,\\
&\frac{d}{dt}\b_1&=\a_2.
\end{array}
$$
It can be then straightforwardly verified that this evolution is determined by the following system of reactions
$$
\begin{array}{lll}
&\b_2^s&\ar\b_2^s,\ \a_1^{-s},\\
&\a_1^s&\ar\a_1^s,\ \b_2^s,\\
&\b_1^s&\ar\b_1^s,\ \a_2^{-s},\\
&\a_2^s&\ar\a_2^s,\ \b_1^s,\ s\in\{ +,-\}.
\end{array}
$$
We can now formulate the following method of building the list of reactions for a Hamiltonian $H$ written through creations and annihilations:
we must multiply $H$ by $-i$ and represent $a_j$ and $a_j^+$ as the annihilation and creation of a.q. taking into account their types ($\a$ for real and $\b$ for imaginary) and their signs. 
This method can be straightforwardly verified in the cases where a Hamiltonian has the form $-\s_y$ and $-\s_z$. 
This method is valid for all one and two particle interactions as well (see simulation of entanglement). The total number of a.q. here grows, and to conserve it we can use the reductions.

\subsection{Representation of unitary evolution by equilibrium reactions}

We now show how to build a list of equilibrium reactions representing a unitary evolution.

Consider sequentially three examples corresponding to the cases
$H=-\s_x,\ -\s_y,\ -\s_z$. 

\nn

{\large \bf First example}
\nn

Let $H=-\sigma_x=-\left(\begin{array}{lll}
&0&1\\
&1&0
\end{array}
\right)$. The action of $U(\delta t)= \left(\begin{array}{lll}
&1&i\delta t\\
&i\delta t&1
\end{array}
\right)$ on the state $\left(\begin{array}{ll}
&\a_1+i\b_1\\
&\a_2+i\b_2
\end{array}
\right)$ results in the state 
\begin{equation}
\left(\begin{array}{ll}
&\a_1-\b_2\delta t+i(\b_1+\a_2\delta t)\\
&\a_2-\b_1\delta t+i(\b_2+\a_1\delta t)
\end{array}
\right).
\end{equation} 
The key observation consists in that $U(\delta t)=L(\Delta t)$ for the appropriate time frame $\Delta t$ and the following list of reactions
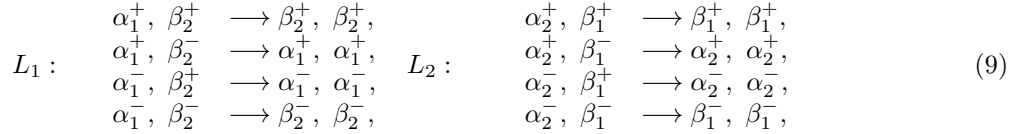
\begin{equation}
L_1:\ \begin{array}{lll}
&\a_1^+,\ \b_2^+&\ar \b_2^+,\ \b_2^+,\\
&\a_1^+,\ \b_2^-&\ar \a_1^+,\ \a_1^+,\\
&\a_1^-,\ \b_2^+&\ar \a_1^-,\ \a_1^-,\\
&\a_1^-,\ \b_2^-&\ar \b_2^-,\ \b_2^-,\\
\end{array}\ \ L_2:\ \ \
\begin{array}{lll}
&\a_2^+,\ \b_1^+&\ar \b_1^+,\ \b_1^+,\\
&\a_2^+,\ \b_1^-&\ar \a_2^+,\ \a_2^+,\\
&\a_2^-,\ \b_1^+&\ar \a_2^-,\ \a_2^-,\\
&\a_2^-,\ \b_1^-&\ar \b_1^-,\ \b_1^-,\\
\end{array} 
\label{L}
\end{equation}
where any a.q. is denoted as the corresponding amplitude $\a_i^s$. 
Indeed, count the preponderance of positive a.q. over negative for all amplitudes $\a_1,\ \b_2$ participating in the part $L_1$ in the time instant $t+\Delta t$:
$[x_i]_{\Delta t} =[x_i^+]_{\Delta t}-[x_i^-]_{\Delta t}$, where $i=1,2$. From the first and second and correspondingly from the third and fourth reactions of $L_1$ we have
\begin{equation}
\begin{array}{lll}
&[\a_1^+]_{\Delta t}&=[\a_1^+]-\g [\a_1^+][\b_2^+]+\g [\a_1^+][\b_2^-],\\
&[\a_1^-]_{\Delta t}&=[\a_1^-]+\g [\a_1^-][\b_2^+]-\g 
[\a_1^-][\b_2^-],
\end{array}
\label{main}
\end{equation}
 where the small parameter $\g$ is the probability of the collision of two fixed a.q. in the time frame $\Delta t$, and $[c]$ denotes the quantity of quanta $c$ in the cirrent time instant $t$. Subtracting these two equations and repeating the corresponding procedure for $\b_2$ we obtain 
\begin{equation}
\begin{array}{lll}
&[\a_1]_{\Delta t}&=[\a_1]-\g [\b_2]\{ \a_1\},\\
&[\b_2]_{\Delta t}&=[\b_2]+\g [\a_1]\{ \b_2\},
\end{array}
\end{equation}
where $\{x_i\}$ denotes the total number of a.q. of this type in the current time instant $t$. 
It results in the system of two differential equations 
\begin{equation}
\begin{array}{lll}
& \frac{d}{dt}[\a_1]&=-\g_0[\b_2] \{ \a_1\},\\
& \frac{d}{dt}[\b_2]&=\g_0[\a_1] \{ \b_2\}.
\label{1}
\end{array}
\end{equation}
Here $\g_0$ denotes the probability of the collision of two fixed a.q. in the unit time frame.
Analogously, adding the equations from (\ref{main}) and doing the same thing for $\b_2$ we obtain the system of equations for total numbers of quanta:
\begin{equation}
\begin{array}{lll}
& \frac{d}{dt}\{ \a_1\}&=-\g_0[\b_2][\a_1],\\
& \frac{d}{dt}\{ \b_2\}&=\g_0[\a_1][\b_2].
\label{2}
\end{array}
\end{equation}
We choose the initial numbers of all quanta so that 
$\g_0$ is very small as well as the initial value of $[x_i]/\{ x_i\}$. Then we have to solve the joint system of differential equations (\ref{1}) and (\ref{2}). An approximate solution can be found if we impose the following additional condition: $\{ x_i\}=A=const$. This condition means simply that we should permanently add some quantity of complementary pairs of the form $(x_i^+,\ x_i^-)$ to the bubble or delete some of them in order to keep unchanged the total number of quanta of the same type\footnote{ It presents no difficulties in computer simulation because the solution of (\ref{1}) with this assumption is clear. It is shown below that this manual intervention to the internal bubble process is unnecessary at all.}. Then the system (\ref{1}) gains the solution of the form 
\begin{equation}
\begin{array}{lll}
&[\a_1]&=-B\cos\w t+C\sin\w t,\\
&[\b_2]&=B\cos\w t+C\sin\w t,
\end{array}
\label{3}
\end{equation}
where $\w=\g_0A$, which is normalized: $[\a_1]^2+[\b_2]^2=const$. We see that the appropriate choice of $\w$ makes this solution equal to the solution of Shroedinger equation with this Hamiltonian. The same calculation can be performed with the part $L_2$ in (\ref{L}) which means that the appropriate choice of the bubble parameters makes the equation (\ref{bubble}) true.  
Considering the case when $A$ depends on the time we mention that its initial value can be made so large that to keep unchanged the initial value of $\w$ we have to choice $\g_0$ very small that in view of (\ref{2}) results in the slow changes in $A$ and gives correspondingly the better approximation of the solution of (\ref{1}),(\ref{2}). That is choosing sequentially values $A_0,A_1,\ldots$ of $A$ converging to infinity, and the corresponding values $\g_0^0,\g_0^1,\ldots$ of $\g_0$ converging to zero, such that $\g_0^n=\w/A_n$, we obtain for each $n$ the solution of (\ref{1}),(\ref{2}) such that $[\a_1],\ [\b_2]$ converge to the solution (\ref{3}).

\nn
{\large \bf Second example}
\nn

Consider the case $H=-\sigma_y=\left(\begin{array}{lll}
&0&i\\
&-i&0
\end{array}
\right)$. Here we have 
$U(\delta t)=1-iH\delta t=
\left(
\begin{array}{lll}
&1&\delta t\\
&-\delta t&1
\end{array}
\right)$. Given a state $(\a_1+i\b_1)|0\rangle+(\a_2+i\b_2)|1\rangle$ this operator transforms it to
$$
\left(\begin{array}{ll}
&\a_1+\a_2\delta t+i(\b_1+\b_2\delta t)\\
&\a_2-\a_1\delta t+i(\b_2-\b_1\delta t)
\end{array}
\right).
$$
Consider the following list of reactions $L_1\cup L_2$:
\begin{equation}
L_1:\ \begin{array}{lll}
&\a_1^+,\ \a_2^+&\ar \a_1^+,\ \a_1^+,\\
&\a_1^-,\ \a_2^+&\ar \a_2^+,\ \a_2^+,\\
&\a_1^+,\ \a_2^-&\ar \a_2^-,\ \a_2^-,\\
&\a_1^-,\ \a_2^-&\ar \a_1^-,\ \a_1^-,\\
\end{array}\ \ L_2:\ 
\begin{array}{lll}
&\b_2^+,\ \b_2^+&\ar \b_1^+,\ \b_1^+,\\
&\b_1^-,\ \b_2^+&\ar \b_2^+,\ \b_2^+,\\
&\b_1^+,\ \b_2^-&\ar \b_2^-,\ \b_2^-,\\
&\b_1^-,\ \b_2^-&\ar \b_1^-,\ \b_1^-,\\
\end{array}
\label{L1}
\end{equation}

The first part of this list can be transformed to the part $L_1$ of the list (\ref{L}) by the following replacement:
$\a_1^+\ar \b_2^+,\ \a_2^+\ar \a_1^+,\ \a_1^-\ar \b_2^-,\
\a_2^-\ar \a_1^-.$ The analogous conclusion takes place for the second part $L_2$ of the list. We thus obtain the corresponding dynamical description of the amplitude quanta as in the first example. 

\nn
{\large \bf Third example}
\nn

Consider the case $H=-\sigma_z=\left(\begin{array}{lll}
-&1&0\\
&0&1
\end{array}
\right)$. Here we have $U(\delta t)=1-iH\delta t=
\left(\begin{array}{lll}
&1+\delta t&0\\
&0&1-\delta t
\end{array}
\right)$. Given a state $(\a_1+i\b_1)|0\rangle+(\a_2+i\b_2)|1\rangle$ this operator transforms it to
$$
\left(\begin{array}{ll}
&\a_1-\b_1\delta t+i(\a_1+\b_1\delta t)\\
&\a_2+\b_2\delta t+i(\b_2-\a_2\delta t)
\end{array}
\right).
$$
Consider the following list of reactions
\begin{equation}
L_1:\ \begin{array}{lll}
&\a_1^+,\ \b_1^+&\ar \b_1^+,\ \b_1^+,\\
&\a_1^-,\ \b_1^+&\ar \a_1^-,\ \a_1^-,\\
&\a_1^+,\ \b_1^-&\ar \a_1^+,\ \a_1^+,\\
&\a_1^-,\ \b_1^-&\ar \b_1^-,\ \b_1^-,\\
\end{array}\ \ L_2:\ 
\begin{array}{lll}
&\a_2^+,\ \b_2^+&\ar \a_2^+,\ \a_2^+,\\
&\a_2^-,\ \b_2^+&\ar \b_2^+,\ \b_2^+,\\
&\a_2^+,\ \b_2^-&\ar \b_2^-,\ \b_2^-,\\
&\a_2^-,\ \b_2^-&\ar \a_2^-,\ \a_2^-,\\
\end{array}
\label{L1}
\end{equation}
It can be reduced to the first example by the corresponding replacement of variables as well. 

We conclude that for the Hamiltonians defined by Pauli matrices (including the identity matrix) the corresponding unitary dynamics can be expressed in the sense of the equality (\ref{bubble}) in terms of reactions of the form $\ldots\ar x,x$. The same conclusion is true for $H\in\{\s_x,\s_y,\s_z\}$. 

If we invert all these reactions in some list $L$ from the previous examples, we obtain the list, representing the inverse Hamiltonian $-H$; we see that a list $L$ can be choosen by many ways.   

\section{Simulation of evolution for arbitrary Hamiltonians}

Take up the case of an arbitrary Hamiltonian. In fact a Hamiltonian reflect the external conditions imposed to the system: the kinetic energy depends on the coordinate system and the potential is determined by the external fields (or by the other particles that are not included to the system). It is then reasonable to assume that the Hamiltonian must be somehow written at the membrane of the bubble and it must influence on a.q. changing their type when they are reflected by the membrane. We expand a Hamiltonian to a sum of Pauli matrices with real nonnegative coefficients $l_{i,j}$ as follows
\begin{equation}
H=\sum\limits_{i,j=0,\ i\leq j}^{N-1}l_{i,j}H_{i,j}
\label{Ham}
\end{equation}
where each $H_{i,j}$ can have nonzero elements among $h_{i,i},\ h_{i,j},\ h_{j,i},\ h_{j,j}$ only and the corresponding part of the matrix belongs to $\{+-I, +-\sigma_x,\ +-\sigma_y,\ +-\sigma_z\}$, $I$ denotes the identity matrix. For any summand taken separately we can express the corresponding unitary dynamics by the list $L_{i,j}$ of reactions. If we divide the set $Q$ of all a.q. in the bubble into the parts $Q_{i,j}$ each of which corresponds to the part $H_{i,j}$ of the Hamiltonian and form the joint list $L_{int}=\bigcup\limits_{i,j=0,\ i\leq j	}^{N-1}L_{i,j}$ of reactions we obtain the list of internal reactions between quanta corresponding to the evolution $U(t)=\exp(-iHt)$. 

We divide the membrane of the bubble into the non-overlapping sets as follows
\begin{equation}
\partial{\cal B}=\bigcup\limits_{i,j=0,\ i\leq j}^{N-1}\partial{\cal B}_{i,j}
\label{division}
\end{equation}

where for every $i,j$-th summand of (\ref{Ham}) the area of the corresponding set ${\cal B}_{i,j}$ is $l_{i,j}$. We supply the lower index of the a.q. type by the auxiliary pair of the form $i,j$ where $i\leq j$, $i,j=0,1,\ldots ,N-1$. Introduce the following rules for collisions with the membrane:
\begin{equation}
\begin{array}{lll}
&x_{i\ \ldots}^s,\ \partial{\cal B}_{i,j}&\ar x_{i\ (i,j)}^s,\ \partial{\cal B}_{i,j}\\
&x_{j\ \ldots}^s,\ \partial{\cal B}_{i,j}&\ar x_{j\ (i,j)}^s,\ \partial{\cal B}_{i,j},
\end{array}
\label{reflection}
\end{equation}
where dots replace any auxiliary option or an empty word. The resulting list $L$ obtained by the addition of all reactions of the form (\ref{reflection}) to $L_{int}$ satisfies the condition (\ref{bubble}).

Alternatively, we could divide the time segment $[0,t]$ into the short time frames of the length $l_{i,j}\Delta t$ each and simulate the actions of all the summands of Hamiltonian on the corresponding time frames in turn. The equality (\ref{bubble}) then results from the above considerations in view of the approximate equation $e^{-iHt}\approx (e^{-iH_{1,1}l_{1,1}\Delta t} e^{-iH_{1,2}l_{1,2}\Delta t}\ldots e^{-iH_{N-1,N-1}l_{N-1,N-1}\Delta t})^{t/\Delta t}$ that follows from the Trotter formula. It can be reached by the list of the form (\ref{reflection}) if we properly change the division (\ref{division}) of the membrane in the corresponding time instances $ l_{1,1}\Delta t,\  (l_{1,1}+l_{1,2})\Delta t,\ \ldots$. The simulation process in any case is thus controlled by the changes of the division (\ref{division})\footnote{It is true for the considered case when we do not represent photons as the separate bubbles. If we do so, then the model will be out of control in-simulation; its evolution will be determined by the initial states of all a.q.} and the list of reactions remains constant and even does not depend on a Hamiltonian. 

The both these methods are valid for the reactions of the type 1) as well as for the more general type 3). 

Consider the lists consisting of reactions of the following form $x^s_j, x^{s'}_{j'}\ar x_j^s,\ x_j^s$, where $j=j_0\ \ldots,\ j'=j_0\ \ldots$, e.g. the reactions touching only a.q. corresponding to the same basic states. Such lists can represent any turn in the Hilbert space of the form $|\Psi\rangle\ar e^{i\phi t}|\Psi\rangle$. Namely, consider a list $L_j$ obtained from the list $L_1$ (see (\ref{L})) by the replacement of $\a_1,\ \b_2$ by the types $\a_j,\ \b_j$ of a.q. corresponding to the amplitude of the $j$-th basic state. Then the list $L=\bigcup\limits_{j=0}^{N-1}L_j$ represents this turn provided we choose the time scale correspondingly to the parameter $\phi$. Denote this list $L$ by $L_{\phi}$. Now we can interpret an eigenvector in terms of reactions. A state $|\Psi\rangle$ is eigenvector of Hamiltonian $H$ corresponding to the frequency $\w$ if and only if the list $L_H$ representing evolution induced by $H$ is equivalent to $L_{\phi}$ relative to the state $|\Psi\rangle$. 

\section{Simulation of classical dynamics as a movement of membrane}

According to our plan the membrane is the single object of the model available to the observation of a user. Hence we can simulate all classical dynamics including measurements of a current quantum state by the membrane movements. We do it such that the bubble occupies the area in space where the probability to find the particle is close to one. Though we refused to associate any physical sense with the movements of a.q. the internal area of the bubble is grained and each its grain with the coordinates of the center $r_j$ represents the basic state of our particle\footnote{Here we suppose here that its spin is zero. For the nonzero spins $s$ we should consider the space of columns of the hight $2s-1$ and our considerations remain true.}. 

In particular, each membrane cell $c$ has the coordinates $r(c)$ which is stored in its type. We denote by $r'(c)$ the coordinates of the grain closest to $c$ and disposed inside the bubble in the line of the normal to the membrane in the point $r(c)$. We choose some threshold value $X_0$ which determines the minimal admissible value of the modulus of amplitude, so that all amplitudes $\la :\ |\la |<X_0$ are considered as zero. For each membrane cell $c$ we get the special meter for the storage of the numbers $[\a ], [\b ]$, where $\a+i\b$ is the amplitude of the state $r'(c)$. As soon as $\sqrt{[\a ]^1+[\b ]^2}$ becomes less than $X_0$, we shift the cell $c$ to the point $r'(c)$ and eliminate all a.q. corresponding to the state $r'(c)$. Analogously, we choose the second threshold value $X_1$ for the modulus of amplitude which we consider as the sufficient to spread the bubble in this direction. As soon as for some membrane cell $c$ the value $\sqrt{[\a ]^1+[\b ]^2}$ becomes greater than $X_1$, we shift this cell in the opposite direction to $r'(c)$, and get the required quantity of the mutually opposite a.q. 
for the new basic state $r(c)$ which comes to be inside the bubble.
In brief, the bubble moves in the line of the modulus of the space state amplitude amplification. We have thus described the classical movement of the bubble corresponding to the current state unitary evolution. 

However, here one significant moment arises that leads to the measurements of the current state. The point is that in the described process of the membrane movements a situation can arise when the bubble loses the connectedness, e.g. transforms to the different bubbles. For example, consider an electron tunneling between two equal potential holes separated by a barrier. The probabilities to find the electron in these two holes will be approximately equal for the states with the lowest energy. It means that the bubble of electron in these states will be separated by the tight bridge in half and the width of this bridge will be the less the higher the barrier is. If we slowly raise the barrier then the moment comes when the bubble splits to two disconnected components. This situation is inadmissible because the number of particles must remain unchanged. Just in the moment when we would tear the membrane to two disconnected pieces the measurement must happen which will choose from these two pieces the one which will represent the electron later on.    
 
Now we describe the procedure of measurement in terms of the "bubble" approach. 
We suppose that a measurement (decoherence)  comes when the membrane ${\cal B}$ acquires an inadmissible shape, e.g. comes to the state preceeding to the separation into two disconnected pieces. The simplest way to represent it is to suppose that some signal spreads to the membrane as the special sign containing in the labels of the membrane cells - we call it a color 1, such that this color is inherited by a.q. colliding with the membrane cells with this color. We could denote this color by the additional lower index of a.q. Reactions corresponding to the a.q. with this color will be all $r$- reductions for $r=0,1,\ldots,N-1$. 
After the time $\Delta t$ sufficient to the completion of these reactions a virtual state of the type $(\ ,\ )$ is introduced to the membrane\footnote{Alternatively, we could suppose that a virtual state is introduced at the beginning of the measurement process when the membrane acquires an inadmissible form - just in the point when it happens.}. When the virtual state acquires the form $(\tau(q),\tau(q))$ the measurement is complete and its result is $l$, where $\tau(q)=x^s_l$, and  $|l\rangle$ is a basic state of the simulated particle. This result determines how the bubble must be rebuilt. Our definitions guarantee that the statistics of measurement will be the same as in case of the direct solution of Shroedinger equation. 

By Erenfest theorem the means of the coordinates and impulses of the quantum particle obey the equations of classical dynamics. The spatial position of the bubble corresponds to the classical approximation of the particle state if we throw off the areas of small probability. The "bubble" method of representation thus takes in quantum as well as classical dynamics.

\section{Simulation of entanglement}

The phenomenon of entanglement between the different particles means that its state belonging to the tensor product of the corresponding one particle spaces cannot be factorized to the two corresponding one particle states.\footnote{ About the features of the simulation of states of ensembles consisting of identical particles - see the next section.} Given a system with $n$ particles $1,2,\ldots,n$ we represent each by its own bubble ${\cal B}_1,{\cal B}_2,\ldots,{\cal B}_n$. To simulate the entanglement between these particles we suppose that these bubbles are in touch with each other, say, as they are enumerated. In case $n=2$ we supply each a.q. by the special identification number (i.n.) such that this number will be inherited only once by the first membrane cell which this a.q. collides with and then will change in collisions with the membrane if we point it out explicitly. In case of $n>2$ i.n. will be inherited as many times as many neighboors the corresponding bubble has. We design i.n. of the cell or a.q. $c$ by $\i(c)$. In what follows we assume $n=2$, the generalization to the other cases is straightforward. 

If two bubbles ${\cal B}_1$ and ${\cal B}_2$ are in touch then their membranes have two areas $A_1$ and $A_2$ correspondingly where each cell $c_1$ from $A_1$ touchs one and only one cell $c_2$ from $A_2$ (and vice versa) which we call associated with $c_1$. No more than two cells can be associated, hence for each i.n. $i$ there is no more than one cell $c_1$ such that $c_1$ is associated with some $c_2$ from $\partial {\cal B}_2$ and $i(c_1)=i$. There is no more than two a.q. $a_1$ and $a_2$ from ${\cal B}_1$ and ${\cal B}_2$ such that the two cells with their i.n. are assosiated. This pair $a_1,a_2$ can be treated as a new a.q. of the state of the system of these two particles; we call these a.q. coupled. Its type $\tau (a_1,a_2)$ is determined uniquelly by the types of $a_1$ and $a_2$. In view of chaotic movement of a.q. before coupling this procedure gives the correct bubble description of the system of two particles. This state after coupling will not be entangled. 

We now consider how to simulate quantum evolutions of many particle systems. All the methods we established above for one particle systems can be extended to many particle systems if we take as a.q. the chains of one particle a.q. of the form:
\begin{equation}
\bar x=x^1,\ x^2,\ldots,x^n,
\label{entangled}
\end{equation}
such that all pairs $x^j,x^{j-1},\ \ j=2,3,\ldots,n$ are coupled. Such chains will be the main object in the many particle simulation. They in fact represent separated states of particles coupled in pairs with the types of their a.q. We assume that two such a.q. of the form (\ref{entangled}) collide with each other or one such a.q. collides with the (joint) membrane if all its corresponding components collide with each other or with the corresponding membranes.

In evolutions induced by one particle Hamiltonians i.n. will be inherited naturally for the reactions of the type 1) and for the reactions of the type 3) from a.q. $\bar x$ by a.q. $\bar z$ in a reaction of the form $(\bar x,\bar y)\ar(\bar z,\bar y)$. It gives the correct transformations of the (joint) bubble induced by one particle Hamiltonians. 
Two particle Hamiltonian $H$ is simulated by the reactions between coupled a.q. of the form $(x^1,x^2),(x'^1,x'^2)\ar (x^1,x^2),(x''^1,x''^2)$ where i.n. are inherited by $x''^1,x''^2$ from $x'^1,x'^2$ correspondingly. The correctness of this definition follows from the linearity of evolution. If $n>2$ this method is true, only for the reactions of the form $(x^1,x^2,\ldots,x^n),(x'^1,x'^2,\ldots,x'^n)\ar (x^1,x^2,\ldots,x^n),(x''^1,x''^2,\ldots,x''^n)$ where the first and the second particle interacts we require that $x'^3=x''^3,\ldots,x'^n=x''^n$ and the basic states corresponding to $x^3,\ldots,x^n$ are the same as the basic states corresponding to $x'^3,\ldots,x'^n$ respectively. 

If we use the secondary quantization for the representation of a two particle Hamiltonian as a sum of summand of the form $b_{j,k,l,m}c^+_jc^+_kc_lc_m$, then the same method is valid as in the one particle case that gives the list of nonequilibrium reactions of the type 4), where new chains (and the corresponding new one particle a.q.) appears in each reaction. 

Let us compare the proposed approach with the conventional one. 
If we solve Shroedinger equation for a system of $n$ particles by some finite-difference method we must consider the states in the joint space 
${\cal H}_{big}={\cal H}^{\bigotimes n}={\cal H}\bigotimes{\cal H}\bigotimes\ldots\bigotimes{\cal H}$, where ${\cal H}$ is the space of states for one particle. Its dimensionality is 
$({\rm dim}\ {\cal H})^n$. In the discrete version it means that we choose a grid in 
${\cal H}$ with some spacing $\d t$, and the basis of the joint space consists of all sequences of the length $n$ of the form $k_1,k_2,\ldots,k_n$, where $k_j$ - is a node of the grid in $j$-th space. We consider a chain of the form (\ref{entangled}). Let $\bar l=(l_1,l_2,\ldots,l_n)$ be the list of basic states corresponding to a.q. $x^1,x^2,\ldots,x^n$. The state $\Psi(t)_{\bar {\cal B}}$, corresponding to the joint bubble then belongs to the linear span  ${\cal L}(t)$ of the set $L(t)$ of all such $\bar l$, for some time instant $t$. The dimensionality of ${\cal L}(t)$ in every time instant $t$ does not exceed the quantity of a.q. in some bubble. Instead 
of the state vector dynamics in the space ${\cal H}_{big}$ of exponential dimensionality we thus will simulate the evolution of the other vector in its subspace ${\cal L}(t)$ of the fixed dimensionality which in turn moves in ${\cal H}_{big}$ with time. 

A decoherence of many particle system can be represented in the framework of our model by two ways.

The first way. A state of one $j$-th particle in the simulated system is defined in the case of entangled state by its density matrix $\rho_j$, which can be computed in our model by the conventional rules. This matrix determines the spatial position of 
$j$-th bubble as in the case of a separated particle. 
Connectedness of a set of bubbles is defined by the touching of their membranes.
If a set of bubbles disintegrates to the different components of connectedness ${\cal B}'_1,\ldots,{\cal B}'_h$, then all coupled i.n. $i,i'$, corresponding to a.q. from the different ${\cal B}'_i$, ${\cal B}'_j$, become not coupled. We thus begin to consider spatially separated parts as independent systems.
This way can be used in the case when one particle moves away from the system, for example, an outgoing electron from atom. 

The second way. A discrete set is called $\e_0$-connected, if it cannot be divided to two components such that the distance between them exceeds $\e_0>0$. The set $L(t_0)$ corresponding to time instant $t_0$, when independent bubbles touch, will be $\e_0$-connected for some $\e_0=O(1/N)$, where the constant depends on $n$, and $N$ is a number of a.q. in one bubble. We choose some $\e_1>\e_0$. If $t_1$ is the maximal time instant for which the set $L(t_1)$ is $\e_1$-connected then in the instant $t_1+1$ it breaks out into several disconnected components. Then we simulate a measurement procedure choosing one component as in the case of one bubble. The bubbles are then rebuilt through the density matrix and the simulation continues. This way generalizes the representation of decoherence in one particle system. It can be applied, for example, in the localization of a rigidly bound many particle system passing through two close slits.

\section{Simulation of identical particles}

The "bubble" approach makes possible to simulate ensembles consisting of identical particles: bosons and fermions. 
Such states must possess the corresponding type of symmetry in permutations of the particles; for bosons a permutation gives the same wavefunction, for fermions it changes the sign. We consider an ensemble consisting of identical particles
$1,2,\ldots,n$, represented by the bubbles ${\cal B}_1,{\cal B}_2,\ldots,{\cal B}_n$ correspondingly which simulate their one particle states. Let $1,2,\ldots,k$ - be the set of all basic states for each particle. We agree, that a.q. in each bubble ${\cal B}_j$ are associated with the cells of its membrane $\partial{\cal B}_j$ as it was described earlier. Our aim is to represent a state of the form $|n_1,n_2,\ldots,n_k\rangle$ in the bubble form, where $n_j$ is the number of particles in the state $j$. We define the procedure of a.q. exchange that will give the required state of the joint system of the bubbles. Consider separately the cases of bosons and fermions.

The case of bosons. We can simulate the states in the Fock space of occupation numbers without resorting to the rebuilding of the membranes because their touch only is important. 
We introduce the list $E$ of all the reactions of the form $x_{j,i}^s,\ y_{l,i}^t\ar y_{l,i}^t,\ x_{j,i}^s$, for all possible types of a.q. $j,l$, belonging to the different bosons with the same i.n. $i$ and $s,t\in \{ +,_\}$. The action of these reactions leads to the exchange of states between the different bosons so that the state $|\Psi\rangle_{{\cal B}_t}$ in the Hilbert space resulted from the action of $E$ in the sufficiently long time frame $t$ will have the symmetrized form corresponding to bosons. 
The one particle states corresponding to this semmetrization can be obtained from the bubble if we eliminate all a.q. but those which correspond to one fixed boson $j$.
It is possible to simulate an action of any Hamiltonian (for example, one particle Hamiltonian) in the joint bubble by using the membrane cells of the corresponding type, adding the list $E$ to all lists of reactions. 

The case of fermions. We agree that the membranes of all bubbles  ${\cal B}_j$ can pass a.q. flying to some sort of cells. Let all the membranes be in touch by pairs - we call the touching cells coupled. Consider some pairs of fermions  $j,l$, $j<l$, which membranes touch by the areas $B_j$ and $B_l$. Let $M_j,M_l$ be coupled membrane cells from $A_j$, $A_l$ correspondingly. We consider the reaction of the form 
$$
x_{j,h,i}^s,\ y_{l,f,i}^r\ \ar_{M_j,M_l}\ y_{j,f,i}^r,\ x_{l,h,i}^{-s},
$$
where $\ar_{M_j,M_l}$ means that two initial a.q. with the same i.n. $i$ fly simultaneously to ${M_j,M_l}$ correspondingly, in the lower index the first part denotes the number of fermion and the second part denotes its basic state.
A.q. trajectories here are arranged as usual so that the exchange of a.q. types takes place with the change of sign in the quanta in the bubble $l$\footnote{Of course, we could ensure the more symmetrical form of the exchange reactions if we use the multiplication on $i$ instead of the change of the sign.}. We consider the exchange list $E$, consisting of all reactions of this form. Analogously to the case of bosons the sufficiently long action of the list $E$ results in the bubble which state corresponds to the antisymmetrized wavefunction of the fermion ensemble. 

\section{Advantages of parallel processing with amplitude quanta}

The main adventage of our approach over the direct solution of Shroedinger equation is that it gives the simple and explicit criteria for the wavefunction reduction in the simulation that makes possible to manage with the limited classical computational resources. In the case of entangled states we operate with chains of a.q. which determine the basic states we consider, when all the other states are excluded from the consideration. It gives the linear growth of the required memory when the number of particle increases. Nevertheless we still can take account of simple entangled states of the simulated system. 
In fact we consider such entangled states only for which the place in the memory is reserved from the very outset in the form of a.q. in the initial bubbles, which i.n. determine all the chains in the future evolution. The total number of chains does not thus exceed the total number of a.q. in the initial bubbles. This method corresponds to the assembling of the more and more complex systems beginning from the elementary particles. We have the explicit criterion for the decoherence as well: it is associated with the break of the bubble to the disconnected parts. It makes possible to  visualize some types of quantum evolution. 

The following advantage of this approach is that it gives an effective way of parallelization of the quantum evolution simulation. Let us parallelize the solution of Shroedinger equation with Hamiltonian $H$ for a system of $k$ processors which are joint together in the net where each of them is connected with no more than the fixed number of others independently of the size of whole net so that we cannot change the geometry of connections. Let at most $k_0\ll k$ from these processors can hang-up. If a processor hangs-up it receives all incoming signals and does not reply. Such a "silent" processor cannot be thus immediately detected and all the others continue their work. It would be natural then to suppose that we can control the work of our processors locally, which means that we can impose the bounds to every connection independently but cannot perform any global control on the computations. At first consider the direct solution of the corresponding Shroedinger equation by the method of finite differences. Try to parallelize the solution by the naive way. Divide the space segment\footnote{By the space we mean the expanded space of the ensemble configurations.} into the equal parts and associate each of them with some processor which will compute the wave function on this part. Each processor then has $2m$ or less neighbors ($m$ is the dimensionality of the configuration space) and interaction provides each of them with the current boundary conditions. This scheme is not resistant to the hang-up of one unknown processor; we thus see that the direct way of quantum behavior simulation meets difficulties when planning the distributed computations. 

Now we apply our "bubble approach". Divide the bubble into areas which are so small that one processor is able to follow the trajectories of all a.q. contained simultaneously in one area and all reactions between them. The states and types of quanta flying into area or flying out of area are transmitted through the corresponding connection between the neighboring processors. We assume that the connection capacity is limited by some constant. Let a small part $\e$ of the processors associated with the internal space of the bubble hang-up. In view of the limitation on the interchange between processors this local termination cannot result in the error exceeding $O(\e )$ for the simulation of the probabilities to obtain the final result and thus for a small $\e$ it cannot break the simulation. The similar argument is valid for the processors which are in charge of the membrane. In the model the spatial positions of the membrane cells are not exactly associated with the real spatial coordinates in the small area occupied by the bubble. The external control over the potential contained in Shroedinger equation can be organized in the same manner by the chaotically moving fictitious particles interacting with the membrane cells. Our approach is thus more robust in parallel mode than the method of finite differences\footnote{In our approach the adventage can be taken from the easy and robust parallelism for inequable Hamiltonians $H(t)$. In such cases the conventional "analytic" approach leads to the calculation of the chronological form $T$ of the exponential in the general formula $U(t)=T\{\exp (-i\int_{t_0}^tH(t)dt)\}$ for the state vector unitary evolution. This calculation requires the peculiar tricks like Vik theorem or methods of partial summing of Feynman diagramms. Such "analytic" methods are very specialized and there is no general way to parallelize them for the numerical computations.}.

\section{Conclusion}

We consider a quantum state as a bubble filled by fictitious particles called amplitude quanta: positive, negative, positive imaginary and negative imaginary quanta - all of the types corresponding to each basic state of the considered ensemble. These quanta are in the chaotic movement. Any quantum unitary evolution is then represented by the chemical like reactions between two close amplitude quanta or between the membrane encircling the bubble and quanta colliding with it. A real observable state of a quantum ensemble appears when two amplitude quanta of the same type collide. It gives the direct substantiation of the probabilistic interpretation of a squared module of amplitude. A measurement is considered as the membrane rebuilding resulted from that the membrane splits to two disconnected pieces.

A physical media is represented as a space filled by bubbles. Movements of the bubbles visible from the outside corresponds to the classical dynamics. Entangled states are simulated by the coupled amplitude quanta in touching bubbles.
Identical particles are simulated trough the exchange of coupled amplitude quanta between the touching bubbles. This approach is convenient to the simulation of complex evolutions where unitary segments alternate with measurements. 

The main adventage of the proposed method over the direct solution of Shroedinger equation is that it makes possible to simulate some class of quantum evolutions with entangled states of many particle quantum systems on classical computers. The next adventages:
\begin{itemize}
\item The model has the natural options which values determine its work and it is free from hidden artifacts like generation of randomness which is not included to the initial state of the system. 
\item Convenience for the presentation of entangled states and identical particles.
\item Convenience for parallelism: simulation in parallel mode resistant to hangs-up in the part of processors.
\end{itemize}

\end{document}